\documentclass[preprint,aps,tightenlines,floatfix,showpacs,twocolumn]{revtex4}
\usepackage{graphics}
\usepackage{bm}
\usepackage{amsmath}
\usepackage{amssymb}
\usepackage{txfonts} 
\usepackage{supertabular}
\usepackage{array}                    

\newcommand{\sfrac}[2]{\textstyle\frac{#1}{#2}}

\begin{document}

\title{An effective two-body model for spectra of clusters 
of $^2{\rm H}$, $^3{\rm H}$, $^3{\rm He}$, and $^4{\rm He}$ with $^4{\rm He}$, 
and $^2{\rm H}$-$^4{\rm He}$ scattering.}

\author{P. R. Fraser$^{1}$}
\email{paul.fraser@curtin.edu.au}
\author{K. Massen-Hane$^{1}$}
\author{A.~S.~Kadyrov$^{1}$}
\author{K. Amos$^{2,3}$}
\author{I.~Bray$^{1}$}
\author{L.~Canton$^{4}$}

\affiliation{
$^{1}$ Department of Physics, Astronomy and Medical Radiation
Sciences, Curtin University, GPO Box U1987, Perth 6845, Australia\\
$^{2}$ School of Physics, University of Melbourne, Victoria 3010, Australia\\
$^{3}$ Department of Physics, University of Johannesburg, P.O. Box 524,
  Auckland Park, 2006, South Africa\\
$^{4}$ $^{4}$ Istituto Nazionale di Fisica Nucleare, Sezione di Padova, I-35131, 
  Italy}

\date{\today}

\begin{abstract}
Four light-mass nuclei are considered by an effective two-body
clusterisation method; $^7$Li as $^3$H$+^4$He, $^7$Be as
$^3$He$+^4$He, $^8$Be as $^4$He$+^4$He, and $^6$Li as
$^2$H$+^4$He. The low-energy spectra of the former three are
determined from single-channel Lippmann-Schwinger equations. For the
latter, two uncoupled sets of equations are considered; those
involving the $^3$S$_1$ and those of the posited $^1$S$_0$ states of
$^2$H. Low-energy elastic scattering cross sections are
calculated from the same $^2$H$+^4$He Hamiltonian, for many angles and
energies for which data are available.  While some of these systems
may be more fully described by many-body theories, this work
establishes that a large amount of data may be explained by these
two-body clusterisations.
\end{abstract}


\maketitle

\section{Introduction}

The scattering and cluster spectra formed by an $\alpha$-particle with
each of the four light mass nuclei considered herein constitute basic
information required for studies of nuclear reactions responsible for
the relative abundances of light atomic nuclei observed throughout the
universe. These arose from the big bang, and in light stars ($\le$ 1.5
M$_\odot$) proton-proton chain reactions lead to the formation of
nuclei up to mass-8. Once the $\alpha$ particles generated in those
reactions are present in sufficient number, the triple-$\alpha$
process can produce ${}^{12}$C; the crucial feature being the energy
of the Hoyle state in ${}^{12}$C lying just above the break-up
threshold.  In the triple $\alpha$-process, the first two
$\alpha$-particles fuse to form $^8$Be whose instability to
$\alpha$-decay results in an equilibrium concentration of ${}^8$Be in
stellar environments.

In recent years, the spectra and elastic scattering of these light
mass cluster systems has become of interest as test beds for modern
theoretical techniques. For example, Refs.~\cite{Qu08,Na11,Hu15} used
a RGM/NCSM method and Ref.~\cite{De06} used a Alt-Grassberger-Sandhas
three-body approach to that end.  An interesting method of analytic
continuation of the elastic scattering data at positive energies to
negative energies was investigated in Ref.~\cite{bl16}, and applied to
extract bound-state properties of the $^2$H+$^4$He system.  Herein we
consider these systems in a much simpler way.  We assume them to be
describable with an effective two-body, single-channel model. None of
the four nuclei forming the clusters have low lying excited states
below nucleon breakup thresholds.  However the compound systems
formed, ${}^6$Li, ${}^7$Li, ${}^7$Be, and ${}^8$Be, do.  We use a
Sturmian expansion approach to solve Lippmann-Schwinger (LS)
equations; an approach that provides a low energy spectrum (bound and
continuum states) of the compound nucleus formed by each cluster
considered, as well as giving the relevant $S$-matrices with which
scattering cross sections can be evaluated.

In this investigation, we calculate the low-energy spectrum of $^6$Li
as the cluster $^4$He$+^2$H, and the low-energy elastic scattering
cross section both using the same interaction potential.
Investigation of the low-energy scattering of deuterons from $^4$He
dates back to experimental work in the 1930s~\cite{Po35}. As noted,
Refs.~\cite{Qu08,Na11,Hu15} used a RGM/NCSM method and considered
$^2$H-$^4$He scattering, amongst other reactions.  Ref.~\cite{De06}
used a Alt-Grassberger-Sandhas three-body method in momentum space at
deuteron energies of 4.81 and 17.0~MeV also for the $^2$H-$^4$He
system.  While details of this scattering may be investigated in a
more fundamental way, e.g. by using three- or six-body approaches, it
remains useful to investigate how much of the spectrum and cross
section may be explained by a simpler two-body clusterization. A
similar model phenomenological semi-microscopic model has been used
recently to calculate phase shifts, for which a good match to data was
obtained~\cite{Du11} as was the calculated $S$-factor for capture.

We have also used the same method to specify the spectra of ${}^7$Li
and ${}^{7,8}$Be from the clusters of $^3$H, $^3$He, and $^4$He with
$^4$He respectively.  Spectra of ${}^7$Li and ${}^7$Be have been found
previously~\cite{Ca06a} by solving the coupled-channel problems of
nucleons coupling to ${}^6$He and ${}^7$Be nuclei allowing for the
nucleons to interact with low excitation states of the nuclei.  The
results agreed well with known states in the spectra. Here we do not
have a coupled-channel problem since, for the range of energies we
consider, all nuclei involved can be taken to be in their ground
states.  The spectra of the two mass-7 nuclei have two bound states
and two resonance states below $\sim$7~MeV excitation.

The last system we consider, ${}^8$Be, has only two resonance states
in its low excitation spectrum, the ground and first excited state at
3.03~MeV. The next resonance state has a centroid of 11.35~MeV.  The
ground state resonance lies just 0.0918~MeV above the two $\alpha$
break-up threshold and is very narrow (5.57 eV); both features
crucially important in the three-$\alpha$ stellar process.  The
$^4$He-$^4$He cluster calculation is of the simplest form in the
effective two-body approach and the two resonance states can be found
with appropriate energy values.

In the next section we give a pr$\acute {\rm e}$cis of the method used
and follow that with a short statement on the forms of charge
distributions used to ascertain the Coulomb interactions of the
clusters. Then in Sec.~\ref{test} we report on the spectra of the
clusters $^7$Li and $^{7,8}$Be found with the method we have used. The
spectra and scattering cross sections for the $^2$H-$^4$He cluster are
then given and discussed in Sec.~\ref{dalpha}.  Conclusions are drawn
in Sec.\ref{concl}.

%%%%%%%%%%%%%%%%%%%%%%%%%%%%%%%%%%%%%%%%%%%%%%%%%%%%%%%%%%%%%%%%%%%%%%%%%%

\section{Sturmian expansion solutions of Lippmann-Schwinger equations}
\label{precis}

The method uses separable expansions of the assumed interaction
potentials between two nuclei.  The form factors in that expansion are
derived from Sturmian functions defined from the chosen two-cluster
interaction potentials. In the cases of $^4$He coupled with
$^3$H, $^3$He, and another $^4$He cluster, the two nuclei have no
excited states of low excitation. For example $^4$He has resonance
states, but they lie above 20~MeV excitation. Thus, we deal with
single channel interactions of a spin-$\frac{1}{2}$ or spin-0 particle
with a spin-0 $^4$He.  With the ${}^2$H-$^4$He clusterisation, we
consider that there are two uncoupled sets of equations to solve;
those formed by the $^3$S$_1$ and, separately, the $^1$S$_0$ states of
the $^2$H.

\begin{widetext}
Then with channels $c = (l,I);J^\pi$, ($l$ the orbital  quantum number
of relative motion, $I$ the spin $0, \frac{1}{2}$ or 1 as appropriate
for the nucleus chosen to cluster with an $^4$He nucleus, the LS
equations for the single channel $T$-matrices have the form, 
%%%%%%%%%%%%%%%%%%%%%%%%%%%%%%%%%%%%%%%%%%%%%%%%%%%%%%%%%%%%%%%%%%%%%% 
\begin{equation}
T_{cc'}^{J^\pi}(p,q;E)\ =\  V_{cc'}^{J^\pi}(p,q)
+ \mu \sum_{c''}
\int_0^\infty V_{cc''}^{J^\pi}(p,x) \frac{x^2}{k^2 - x^2 + i\epsilon}
T_{c''c'}^{J^\pi}(x,q;E)\ dx ,
\label{multiTeq}
\end{equation}
%%%%%%%%%%%%%%%%%%%%%%%%%%%%%%%%%%%%%%%%%%%%%%%%%%%%%%%%%%%%%%%%%%%%%% 
where the momentum $k$ = $\sqrt{\mu E}$, with
$\mu$ designating $2m_{\rm red}/\hbar^2$; $m_{\rm red}$ being the reduced mass.  
Solutions of Eq.~(\ref{multiTeq}) are sought using the (finite sum) expansion 
%%%%%%%%%%%%%%%%%%%%%%%%%%%%%%%%%%%%%%%%%%%%%%%%%%%%%%%%%%%%%%%%%%%%%% 
\begin{equation}
V_{cc'}(p,q) \sim  \sum^N_{n = 1} {\hat \chi}_{cn}(p)\
\eta^{-1}_n\ {\hat \chi}_{c'n}(q)\ .
\label{finiteS}
\end{equation}
%%%%%%%%%%%%%%%%%%%%%%%%%%%%%%%%%%%%%%%%%%%%%%%%%%%%%%%%%%%%%%%%%%%%%% 
To evaluate scattering cross sections, one needs the $S$-matrices
which are linked to the $T$-matrices as~\cite{Ca91,Pi95}
%%%%%%%%%%%%%%%%%%%%%%%%%%%%%%%%%%%%%%%%%%%%%%%%%%%%%%%%%%%%%%%%%%%%%% 
\begin{equation}
S_{cc'}\ =\ \delta_{cc'}\ -\ i \pi \mu\ k_{c'}\ T_{cc'}
=\ \delta_{cc'}\ -\ i^{\left(l_{c'} - l_c +1\right)} 
\pi \mu \  \sum_{n,n' = 1}^N \hspace{-2mm}\sqrt{k_c} \;  
{\hat \chi}_{cn}(k_c)\ \left([\mbox{\boldmath $\eta$} - {\bf G}_0]^{-1}
\right)_{nn'}\ {\hat \chi}_{c'n'}(k_{c'}) \sqrt{k_{c'}} \ ,
\label{multiS}
\end{equation}
%%%%%%%%%%%%%%%%%%%%%%%%%%%%%%%%%%%%%%%%%%%%%%%%%%%%%%%%%%%%%%%%%%%%%% 
In this representation, \textbf{${{\bf G}_0}$} 
and \mbox{\boldmath $\eta$} have matrix elements 
%%%%%%%%%%%%%%%%%%%%%%%%%%%%%%%%%%%%%%%%%%%%%%%%%%%%%%%%%%%%%%%%%%%%%% 
\begin{equation}
\left[{\bf{G}}_0 \right]_{nn'} =\ \mu \sum_c \int_0^\infty
{\hat \chi}_{cn}(x) \frac{x^2}{k^2 - x^2 + i\epsilon} {\hat 
\chi}_{cn'}(x)\ dx \hspace*{0.5cm};\hspace*{0.5cm} 
\left[{\mbox {\boldmath $\eta$ }}\right]_{nn'} =\ \eta_n\ \delta_{nn'} \, .
\label{xiGels}
\end{equation}
%%%%%%%%%%%%%%%%%%%%%%%%%%%%%%%%%%%%%%%%%%%%%%%%%%%%%%%%%%%%%%%%%%%%%% 
Bound states of the compound system, if they exist, are defined by the
zeros of the matrix determinant in Eq.~(\ref{multiS}) when the energy,
$E$, is less than zero.

The input matrices of potentials are taken to have the form
%%%%%%%%%%%%%%%%%%%%%%%%%%%%%%%%%%%%%%%%%%%%%%%%%%%%%%%%%%%%%%%%%%%%%% 
\begin{equation}
V_{cc'}(r) =  V_{cc'}^{\rm coul}(r) +
\bigg[ V_0\ \delta_{c'c}\ f(r)\ +\ V_{\ell \ell}\ f(r)\ 
[ {\bf {\ell \cdot \ell}} ]  
\ +\  V_{II}\ f(r)\ [{\bf I \cdot I}] 
\ +\  V_{\ell I}\ g(r)\ [ {\bf {\ell \cdot I}} ]\bigg]_{cc'}
\label{www1}
\end{equation}
%%%%%%%%%%%%%%%%%%%%%%%%%%%%%%%%%%%%%%%%%%%%%%%%%%%%%%%%%%%%%%%%%%%%%% 
wherein local form factors (Woods-Saxon functions), 
%%%%%%%%%%%%%%%%%%%%%%%%%%%%%%%%%%%%%%%%%%%%%%%%%%%%%%%%%%%%%%%%%%%%%% 
\begin{equation}
f(r) = \left[1 + e^{\left( \frac{r-R}{a} \right)} \right]^{-1}
\hspace*{0.3cm} ; \hspace*{0.3cm} g(r) = \frac{1}{r} \frac{df(r)}{dr} ,
\label{radforms}
\end{equation}
%%%%%%%%%%%%%%%%%%%%%%%%%%%%%%%%%%%%%%%%%%%%%%%%%%%%%%%%%%%%%%%%%%%%%% 
are used. If needed, the surface can be deformed ($R = R(\theta \phi)
= R_0\left[1 + \epsilon \right]$).  Details of this and of the
relevant matrix elements are given in Ref.~\cite{Am03}.  $V^{\rm
  coul}_{cc'}(r)$ are elements of the Coulomb potential matrix.  The
forms we use are given in the next section.

\end{widetext}

%%%%%%%%%%%%%%%%%%%%%%%%%%%%%%%%%%%%%%%%%
\section{Charge distributions for the nuclei and the Coulomb interaction between them.}
\label{coulfn}

We assume both nuclei in the cluster have finite charge distributions of
three parameter Fermi (3pF) form, {\it viz}
%%%%%%%%%%%%%%%%%%%%%%%%%%%%%%%%%%%%%%%%%%%%%%%%%%%%%%%%%%%%%%%%%%%%%%%%
\begin{equation}
\rho_{ch}(r) = \rho_0 \left[ 1 + w_c \left( \frac{r}{R_c} \right)^2 \right]
\frac{1}{1 + \exp\left(\frac{r - R_c}{a_c} \right)} ,
\end{equation}
%%%%%%%%%%%%%%%%%%%%%%%%%%%%%%%%%%%%%%%%%%%%%%%%%%%%%%%%%%%%%%%%%%%%%%%%
where $R_c$ and $a_c$ are the radius and diffuseness parameters
for a Woods-Saxon distribution, and $w_c$ is a scaling parameter.
The central charge density is that with which the volume integral
of the distribution equates to the charge of the nucleus represented.
%%%%%%%%%%%%%%%%%%%%%%%%%%%%%%%%%%%%%%%%%%%%%%%%%%%%

\begin{figure}[h]
\scalebox{0.43}{\includegraphics*{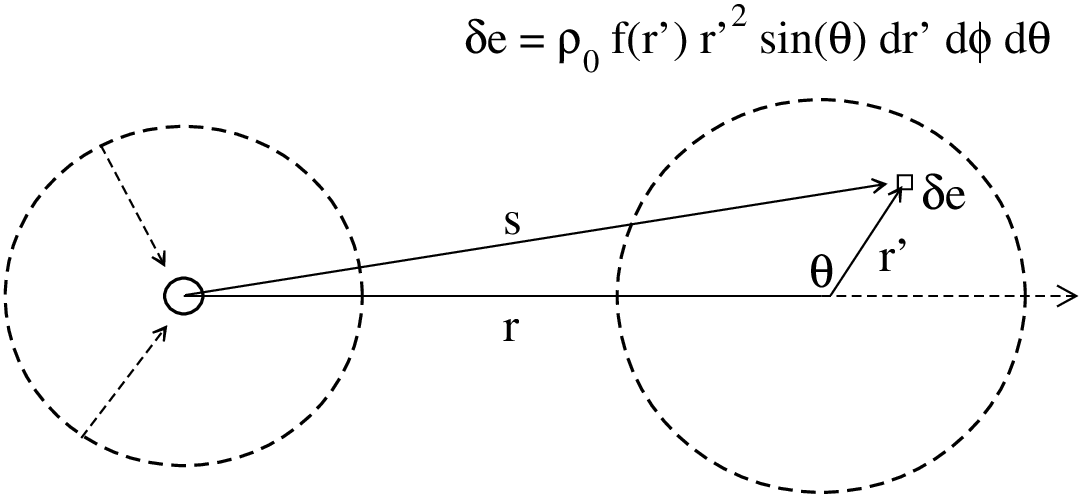}}
\caption{\label{Fig1}
The geometry for two interacting nuclei, both having a 3pF charge distribution.}
\end{figure}

To define the Coulomb interaction between such charge distributions,
first consider that felt by a positively-charged point
test particle with charge $\delta e$ and a general spherical
charge distribution, $\rho_0 f(r)$, i.e.  
%%%%%%%%%%%%%%%%%%%%%%%%%%%%%%%%%%%%%%%%%%%%%%%%%%%%%%
\begin{equation}
V^{(pt)}_{\rm coul}(r) = \delta e \int \rho_0 f(r^\prime) 
\frac{1}{|{\mathbf r^\prime} - {\mathbf r}|}
d{\mathbf r^\prime} \;.
\end{equation}
%%%%%%%%%%%%%%%%%%%%%%%%%%%%%%%%%%%%%%%%%%%%%%%%%%%%%%
After expanding in multipoles and performing angular integration, the
only non-zero component comes from the $s$-wave ($\ell = 0$), whence
%%%%%%%%%%%%%%%%%%%%%%%%%%%%%%%%%%%%%%%%%%%%%%%%%%%%
\begin{equation}
V^{(pt)}_{\rm coul}(r) = 4 \pi (\delta e) \rho_0 \int_0^\infty f(r^\prime) 
v_{\ell = 0}(r^\prime, r) {r^\prime}^2 dr^\prime .
\end{equation}
%%%%%%%%%%%%%%%%%%%%%%%%%%%%%%%%%%%%%%%%%%%%%%%%%
where $v_{\ell =0}(r^\prime, r) = \frac{1}{r_>}$ with $r_>$ and $r_<$
being the greater and lesser of $r^\prime$ and $r$, respectively.
The radial integration splits into two terms, giving
%%%%%%%%%%%%%%%%%%%%%%%%%%%%%%%%%%%%%%%%%%%%%%%%%%%%%%%%%%%%%%%%%%%%%%
\begin{align}
V^{(pt)}_{\rm coul}(r) &= 4 \pi (\delta e) \rho_0 
\left[ \frac{1}{r} \int_0^r f(s)\ s^2\ ds\ \right. 
\nonumber\\
&\hspace{2cm}+\ \left.
\int_r^\infty \frac{1}{s} f(s)\ s^2\ ds \right] \;.
\label{pt-rho}
\end{align}
%%%%%%%%%%%%%%%%%%%%%%%%%%%%%%%%%%%%%%%%%%%%%%%%%%%%%%%%%%%%%%%%%%%%%%%%
With both nuclei in the clusterisation having 3pF charge
distributions, the field given in Eq.(\ref{pt-rho}) is folded with the
3pF charge distribution for the second body.  The geometry is as shown
in Fig.~\ref{Fig1}.
%%%%%%%%%%%%%%%%%%%%%%%%%%%%%%%%%%%%%%%%%%%%%%%%%%%%%%%%%%
%%%%%%%%%%%%%%%%%%%%%%%%%%%%%%%%%%%%%%%%%%%%%%%%%%%%%%%%%%%%%%%%%%%%%%

With $s = \sqrt{r^2 + {r'}^2 - 2 r r^\prime \cos(\theta) }$,
the Coulomb potential is
%%%%%%%%%%%%%%%%%%%%%%%%%%%%%%%%%%%%%%%%%%%%%%%%
\begin{equation}
V_{\rm coul}(r) = 2\pi \hspace{-1.5mm}\int_0^\infty \hspace{-2mm}{r^\prime}^2 f(r^\prime) dr^\prime \hspace{-1mm}
\int_0^\pi \hspace{-2mm}V^{(pt)}_{\rm coul}(s)\ \sin(\theta)\ d\theta .
\end{equation}
%%%%%%%%%%%%%%%%%%%%%%%%%%%%%%%%%%%%%%%%%%%%%%%%%%%%

For ${}^4$He, the parameter values of the 3pF charge distribution are
as given in Ref.~\cite{Vr74,Vr87}.  They are $R_c = 1.008$~fm, $a_c =
0.327$~fm, and $w$~=~0.445.  As ${}^3$H is listed~\cite{Vr74,Vr87} as
also having a root-mean-square (rms) charge radius of 1.7~fm, the
${}^4$He parameter set has been used for its charge distribution as
well. ${}^3$He is listed~\cite{Vr74,Vr87} as having a slightly larger
rms charge radius, 1.88~fm. As there is no specified set of 3pF
parameters given, we considered a range of values for them, since, as
shown in Ref.~\cite{Fr16}, variation in the three parameters leads to
minimal difference in results provided the rms charge radius is kept
constant.  The set used are listed in Table~\ref{He3-3pF}.
%%%%%%%%%%%%%%%%%%%%%%%%%%%%%%%%%%%%%%%%%%%%%%%%%%%%%%%%%%%%%%%%%%%%%%%%%%
\begin{table*}[ht]
\caption{\label{He3-3pF}
Diverse 3pF parameter values giving a root-mean-square 
charge radius of 1.88~fm.} 
\begin{ruledtabular}
\begin{tabular}{l|ccccccccc}
$R_c$ & 1.02 & 1.02 & 1.04 & 1.04 & 1.06 & 1.06 & 1.08 & 1.08 & 1.1 \\
$a_c$ & 0.358 & 0.362 & 0.358 & 0.362 & 0.356 & 0.36 & 0.356 & 0.36 &
0.356\\
$w$ & 0.49 & 0.43 & 0.48 & 0.42 & 0.5 & 0.44 & 0.49 & 0.46 & 0.48
\end{tabular}
\end{ruledtabular}
\end{table*}
%%%%%%%%%%%%%%%%%%%%%%%%%%%%%%%%%%%%%%%%%%%%%%%%

For $^2$H, the rms charge radius has been determined~\cite{Ga01} to be
2.13~fm. To have that value with the average distribution of the
single proton smeared out over an appreciable distance, that rms
radius is met using the set of 3pF parameters, $R = 0.012$~fm, $a_c =
0.592$~fm, and $w_c =0$.

%%%%%%%%%%%%%%%%%%%%%%%%%%%%%%%%%%%%%%%
\section{Studies of the ${}^3{\rm H}+{}^4{\rm He}$, 
${}^3{\rm He}+{}^4{\rm He}$ and ${}^4{\rm He}+{}^4{\rm He}$
systems}
\label{test}

These cases are taken to be single channel problems given that the
components are quite strongly bound and have no excited states below
nucleon emission thresholds. However, the compound systems do have well
established spectra and, for the $^3$H+$^4$He and $^3$He+$^4$He
systems, the states that we might expect to obtain with a potential
model are those indicated in Table~\ref{7Li-states}. The reactions
involving $^4$He that lead to them, or have the mass-7 states as a
compound system, are indicated by the check marks.
%%%%%%%%%%%%%%%%%%%%%%%%%%%%%%%%%%%%%%%%%%%%%%
\begin{table*}[ht]
\caption{\label{7Li-states} 
  States in  ${}^{7}$Li and of ${}^7$Be relevant to this investigation
  and known reactions~\cite{Fi96}  involving $^4$He that populate them.}
\begin{ruledtabular}
\begin{tabular}{l|ccc|ccc}
J$^\pi$ & & ${}^7$Li & & & ${}^7$Be & \\
\hline
$J^\pi$ & $^3$H$(^4$He$,$n$)$ & $^4$He$(^3$He$,\pi^+)$ & $^4$He$(^4$He,$p)$
& $^4$He$(^3$He$,\gamma)$ & $^4$He$(^3$He$,^3$He$),(^3$He$,$p$)$ & $^4$He$(^4$He$,$n$)$\\
\hline
$\sfrac{3}{2}^-$ & $\surd$  & $\surd$ & $\surd$ 
& $\surd$ &   & $\surd$ \\
$\sfrac{1}{2}^-$ &   & $\surd$ & $\surd$ 
& $\surd$ &   & $\surd$ \\
$\sfrac{7}{2}^-$ &   &   & $\surd$ 
&   & $\surd$ & \\
$\sfrac{5}{2}^-$ &   & $\surd$ &   
&   & $\surd$ & \\
\hline
\end{tabular}
\end{ruledtabular}
\end{table*}
%%%%%%%%%%%%%%%%%%%%%%%%%%%%%%%%%%%%%%%%%%%%%%%%

No orthogonalizing pseudo-potential (OPP)~\cite{Am13} to effect
inclusion of the Pauli principle has been used in treating these
clusters as single-channel problems since all states found thereby are
orthogonal.  Thus any state that should be blocked because it requires
the 7 or 8 nucleons to lie in the $0s$-shell simply can be ignored.
Only if there is channel coupling does a problem arise in ensuring
that the Pauli principle is satisfied~\cite{Am03}. With channel
coupling, all resultant states of the cluster are linear combinations
of all states of the same spin-parity defined in the potentials for
each of the target states considered.

\subsection{The ${}^3{\rm H}+{}^4{\rm He}$ and ${}^3{\rm He}+ {}^4{\rm He}$ systems}

Spectra of ${}^7$Li and ${}^7$Be have been found
previously~\cite{Ca06a} using the multi-channel algebraic scattering
(MCAS) program written for spin-$\frac{1}{2}$ particles coupling to a
nucleus. The results agreed well with known states in the spectra. A
program has now been written for $^4$He (spin-0) particles
coupling to a nucleus.  This has been used to again calculate the
spectra for the compound nuclei, ${}^7$Li and ${}^7$Be, as a check against
the results found earlier~\cite{Ca06a}.
%%%%%%%%%%%%%%%%%%%%%%%%%%%%%%%%%%%%%%%%%%%%%%%%%%%%%%%%%%%%%%%
\begin{table*}[ht]
\caption{\label{mass7}
Spectra of ${}^7$Li and ${}^7$Be from a $^4$He
coupled to ${}^3$H and ${}^3$He respectively.
The energies are in MeV while the widths are in keV.
The experimental values are those listed in Ref.~\cite{Ti02}.}
\begin{ruledtabular}
\begin{tabular}{c|cccc|cccc}
 & ${}^7$Li & & & & ${}^7$Be & \\
$J^\pi$ & Exp. & present & check & Ref.~\cite{Ca06a} &
 Exp. & present & check & Ref.~\cite{Ca06a}\\
\hline
$\frac{3}{2}^-$ & spurious  & $-$31.1 & $-$29.6 & $-$29.4  
& spurious & $-$29.7 & $-$27.8 & $-$28.0 \\
$\frac{1}{2}^-$ & spurious  & $-$29.6 & $-$28.0 & $-$27.8
& spurious & $-$28.3 & $-$26.3 & $-$26.4 \\
\hline
$\frac{3}{2}^-$ & $-$2.47 & $-$2.49 & $-$2.59 & $-$2.47
& $-$1.59 & $-$1.55 & $-$1.53 & $-$1.53\\
$\frac{1}{2}^-$ & $-$1.99 & $-$1.81 & $-$1.87 & $-$1.75
& $-$1.16 & $-$0.90  & $-$0.85 & $-$0.84 \\
$\frac{7}{2}^-$ & 2.18\ (69) & 2.23\ (83) & 2.09\ (80) & 2.12\ (83)
& 2.98\ (175) & 3.19\ (180)& 3.14\ (204) & 3.07 (180)\\
$\frac{5}{2}^-$ & 4.13\ (918) & 4.16\ (717) & 4.05\ (800) & 4.12\ (834)
& 5.14\ (1200) & 5.15 (1040) & 5.13\ (1250) & 5.09\ (1194)\\
\end{tabular}
\end{ruledtabular}
\end{table*}
%%%%%%%%%%%%%%%%%%%%%%%%%%%%%%%%%%%%%%%%%%%%%%%%%%%%%%%%%%%%

For the check run, the interaction with strength parameter values (in
MeV), $V_0$ = -76.8, $V_{l l}$ = 1.15, and $V_{l I}$ = 2.34 was used.
The geometry of the Woods-Saxon form was set with $R_0$ = 2.39 and $a$
= 0.68~fm. The Coulomb potential was set, as in Ref.~\cite{Ca06a}, to be
that from a uniformly charged sphere. The charge radius for the
${}^4$He+${}^3$H calculation was taken as $R_c$ = 2.34~fm, while a
slightly larger charge radius (2.39) was used for the $^3$He+$^4$He
calculation.  These values differ (slightly) from those used
previously~\cite{Ca06a} in a study of the same compound systems but
taken as ${}^3$H and ${}^3$He projectiles coupled to an $^4$He target. The
differences are due primarily to our current use of the nuclear masses
listed in Ref.~\cite{Au03} rather than the nucleon mass numbers.  Using
this interaction, we obtained the results listed in Table~\ref{mass7}
and in the columns with the heading `check'.  The comparison between
the results given in Ref.~\cite{Ca06a} and by these check runs is
sufficiently good that the two codes used we deem to give equivalent
results.

Using 3pF distributions for both nuclei in the clusters instead of the
uniform sphere approach above, and with adjusted nuclear potential
parameter values, the results listed in Table~\ref{mass7} in the
columns specified as 'present' were obtained.  For these results, the
nuclear interaction parameter values were $V_0$~=~$-$80.15~MeV,
$V_{\ell\ell}$~=~1.1~MeV, and $V_{\ell I}$~=~3.0~MeV with a
Woods-Saxon geometry, $R_0$~=~2.35~fm and $a_0$~=~0.64~fm. The 3pF
parameter set defined above to give an rms charge radius of 1.7~fm was
used for both ${}^3$H and ${}^4$He, while that used for ${}^3$He we
choose to be the first set in Table~\ref{He3-3pF}, namely $R_c$~=~1.02~fm, 
$a_c$~=~0.358~fm, and $w_c$~=~0.49.  Using the other sets of
parameter values listed in Table~\ref{He3-3pF} (all of which gave an
rms charge radius of 1.88 fm) varied the spectral energies from those
listed by no more 25~keV (centroids and widths).
 
The `present' results agree to within 200~keV (energies and widths).
This is encouraging  since only the $^4$He break-up thresholds
(2.47 and 1.59~MeV for  ${}^7$Li and ${}^7$Be) lie in the range shown.

\subsection{The ${}^4{\rm He}+{}^4{\rm He}$ system}

We have evaluated the spectrum resulting for the clusters
$^4$He+$^4$He as another single-channel problem, since the $^4$He
nucleus is strongly bound and has no other bound state in the
(low-energy) spectrum.  From Ref.~\cite{Fi96}, we note that the
$0_1^+$ and $2_1^+$ states of ${}^8$Be have been found with the
$^4$He$(^4$He$,\gamma)$ and $^4$He$(^4$He$,^4$He) reactions.  With a
(positive-parity) interaction [$V_0 = -47.1$~MeV, $V_{ll} = 0.4$~MeV,
  $R_0 = 2.1$~fm, and $a_0 = 0.6$~fm] and the Coulomb potential from
folding two 3pF distributions, two low-excitation resonance states for
${}^8$Be, relative to the cluster threshold, are found. They are the
ground state ($0^+$) resonance having centroid and width energies of
0.092~MeV and 5 eV [c/f experimental values\cite{Ti02} 0.092~MeV and
  5.96~eV] and a first excited ($2^+$) resonance state with centroid
and width energies of 3.16~MeV and 1.11~MeV compared with experimental
values of 3.03~MeV and 1.51~MeV respectively.  With this simple (local
Woods-Saxon) single-channel interaction, no $4^+$ resonance state is
found; at least below 20~MeV excitation.

In this case, the interaction allows a $0s$-state bound by
20~MeV, which, due to Pauli blocking, is deemed to be spurious and
so has been ignored since all resultant states from the single channel
problem are orthonormal.

\section{Results for the ${}^2{\rm H}+{}^4{\rm He}$ system; spectrum of 
$^6{\rm Li}$ and scattering.}
\label{dalpha}

We consider the $^2$H-$^4$He system as two single-channel problems;
one for the $^3$S$_1$ (ground) state and the other for the posited $^1$S$_0$
state of the deuteron.  We do not consider the states to be coupled by
a spin-isospin changing interaction.  The deuteron states are both of
positive parity and the low excitation spectrum of ${}^6$Li only has
positive parity states so the dominant character of the interaction
potentials is of positive parity.  The results were obtained using
$V_0$ = -64.775, $V_{ll}$ = 0.93, $V_{lI}$ = 1.97, and $V_{II}^+$ =
-2.0 (all in MeV) with a geometry of $R_0$ = 2.3 and $a_0$ = 0.43~fm.
We also allowed the potential to have second order deformation
contribution with $\beta_2$ = 0.22.  No negative parity interaction
has been used, as no such states are known.

$^6$Li has a known low-energy spectrum containing six states: a
$1^+;0$ ground state, followed by $3^+;0$, $0^+;1$, $2^+;0$, $2^+;1$
states, and finally a second $1^+;0$ at 5.65~MeV. The next state is
17.98~MeV above the ground state.  The $3^+$ state appears as a clear
resonance in the $^2$H+$^4$He cross section, 2.186~MeV above the
ground state (or 0.7117~MeV above the scattering threshold, at $E_d =
1.067$~MeV or $E_\alpha = 2.135$
MeV)~\cite{La53,Ga55,Ba79,Na85,Be86,Ke93,Qu93}.  Also evident is the
$2^+_1$ resonance 4.31~MeV above the $^6$Li ground state (or 2.8375
MeV above the scattering threshold, at $E_d = 4.253$~MeV or $E_\alpha
= 8.507$~MeV)~\cite{Ga55,Oh64,Br82,Is04}. Present but less pronounced
is the $1^+$ resonance 5.65~MeV above the $^6$Li ground state (or
4.1757~MeV above the scattering threshold at $E_d = 6.264$~MeV or
$E_\alpha = 12.527$)~\cite{Ma68,Ha77}. It is possible that
Ref.~\cite{Se64} shows data for the $0^+$ resonance of $^6$Li 3.563
MeV above the ground state (or 2.0887~MeV above the scattering
threshold at $E_d = 3.133$~MeV or $E_\alpha = 6.266$~MeV), but the
data points are sparse. The $2^+_2$ state of $^6$Li 4.31~MeV above the
ground state (or 2.8357~MeV above the scattering threshold) does not
appear in data. Data also exists for higher
energies~\cite{Al51,Fr54,Oe63,Ka78,Gr75,St80}.

\begin{figure*}[htp]
\begin{center}
\scalebox{0.7}{\includegraphics*{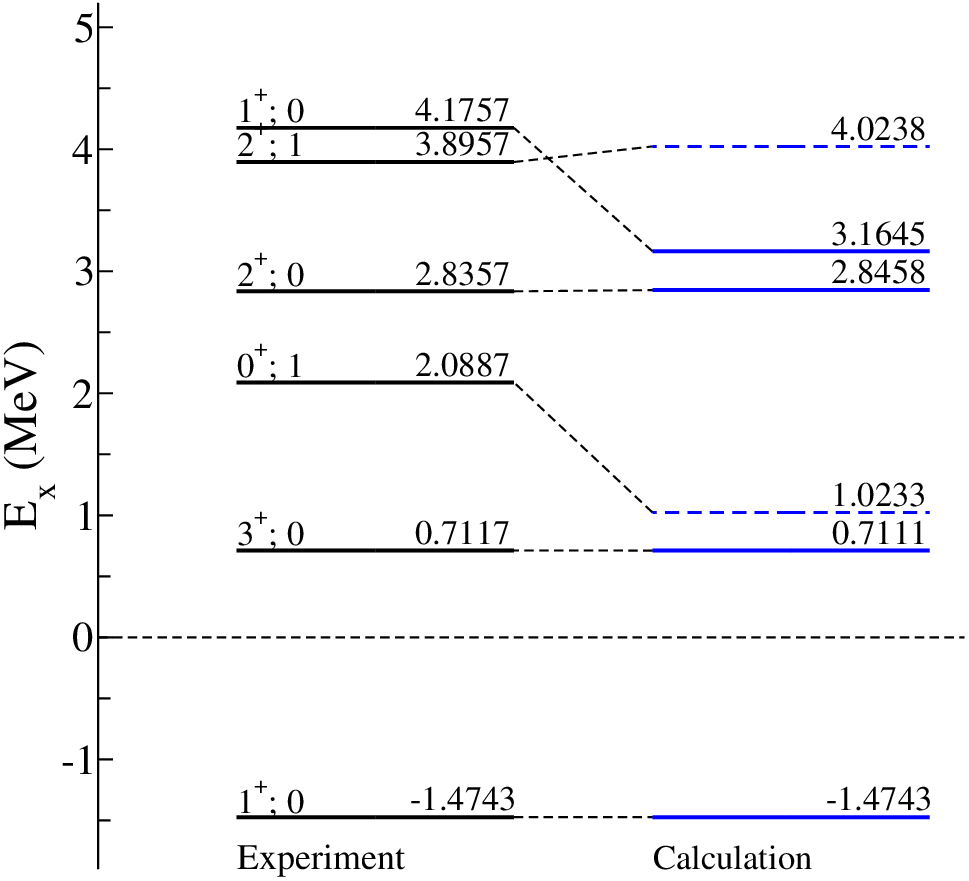}}
\end{center}
\caption{ \label{Fig2}(Color online.) Experimental spectrum of $^6$Li
  compared with the present calculation. In the calculation, solid
  lines are from coupling to the deuteron $^3$S$_1$ state, and dashed
  lines are from coupling to the $^1$S$_0$ state.}
\end{figure*}

In Fig.~\ref{Fig2} the experimentally known spectrum is compared with that
resulting from the calculation.
%%%%%%%%%%%%%%%%%%%%%%%%%%%%%%%%%%%%%%%%%%%%%%%%%%%%%%%%%%%%%%%%%%%%%%%%%%%%%%%
%%%%%%%%%%%%%%%%%%%%%%%%%%%%%%%%%%%%%%%%%%%%%%%%%%%%%%%%%%%%%%%%%%%%%%%%%%%%%%%
The calculation finds all six known low-energy states of $^6$Li.
Spurious minimal energy states were eliminated when an OPP
contribution of $\lambda = 10^6$~MeV was used to block the
$1s_{\frac{1}{2}}$ single-nucleon orbit from having more than the four allowed
nucleons. They can also simply be discarded, since they are orthogonal
to the all others.

Owing to the absence of coupling between channel involving the the
$^2$H triplet and singlet states, the $0^+_1$ and $2^+_1$ states are
purely found from coupling of the $^2$H singlet state to the $^4$He
ground state partial waves. All other states are purely found from
coupling of the deuteron triplet state to the $^4$He ground state. The
first three excited states are found to within a few tens of eV. The
final $T=0$ state, the $1^+_2$, is too low in energy by an MeV.  The
singlet state was assumed to be at the $^2$H breakup threshold, i.e.,
2.224~MeV above the ground state.  As there is no mixing between the
$^6$Li $T=0$ and $T=1$ states in this calculation, the excitation
energies of the two $T=1$ states depend linearly on the energy of the
$^2$H singlet state, though the gap between them is set by the
interaction potential parameters. This gap is too large by
$\sim$1.1~MeV, and while the energy of the $2^+_2$ state is recreated
well, the calculated energy of the $0^+$ state is too low. It is
likely that the antibound singlet state would have a different charge
distribution and a different nuclear interaction with the
$\alpha$-particle than the triplet state. However, in this work we opt
to use a single interaction as experimental data is not available to
guide selection of the relevant parameters.

\begin{figure*}[htp]
\begin{center}
\scalebox{0.83}{\includegraphics*{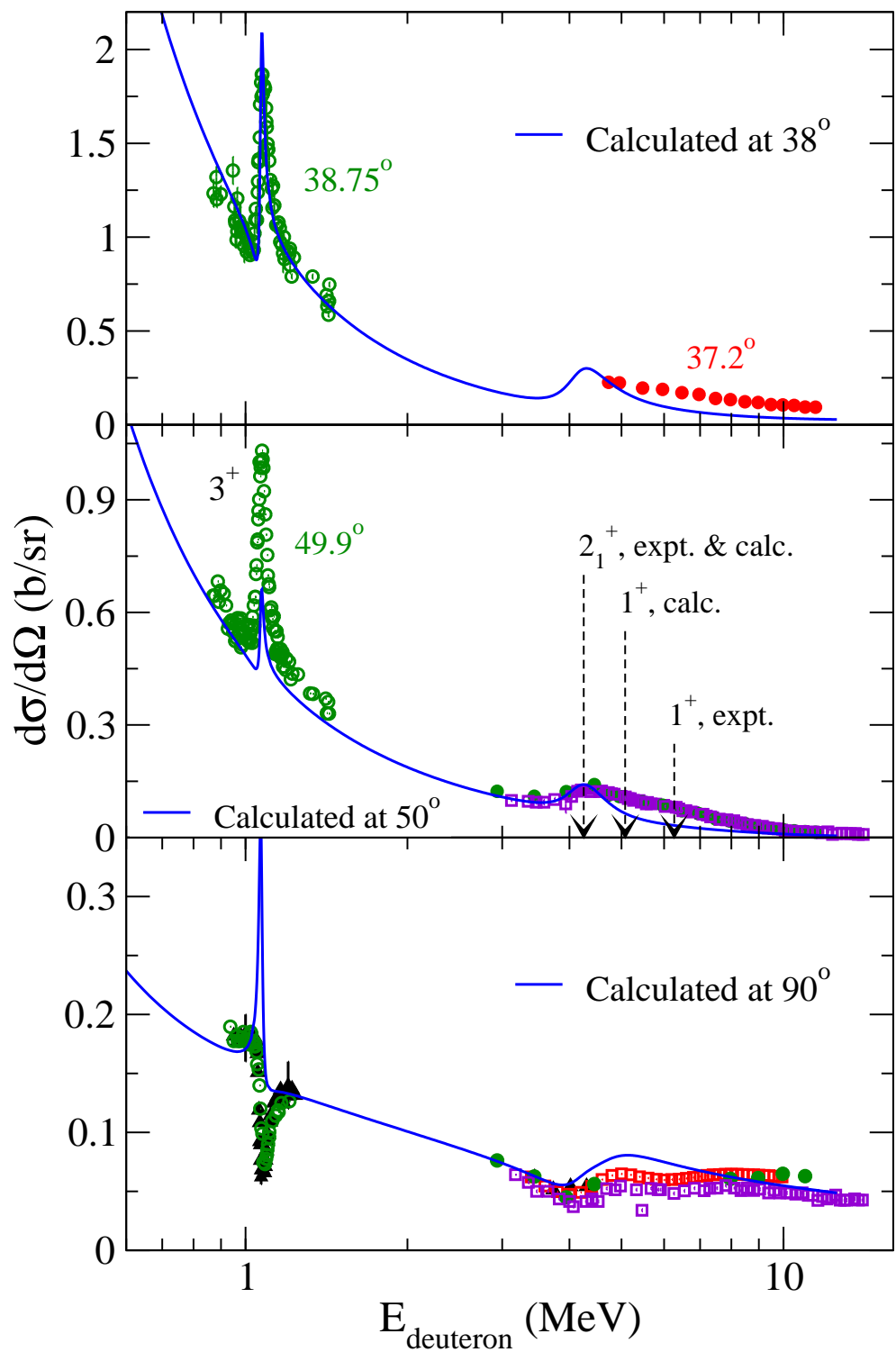}}
\end{center}
\caption{ \label{Fig3}(Color online.) Experimental elastic cross
  sections for $^2$H+$^4$He scattering, at fixed angles, compared with
  the calculations. The data are from
  Refs.~\cite{La53,Ga55,Oh64,Se64,Ma68,Ba79}.  The 50$^\circ$ panel
  shows energies where resonances are found in the spectrum, both observed
  and calculated.}
\end{figure*}

\begin{figure*}[htp]
\begin{center}
\scalebox{0.83}{\includegraphics*{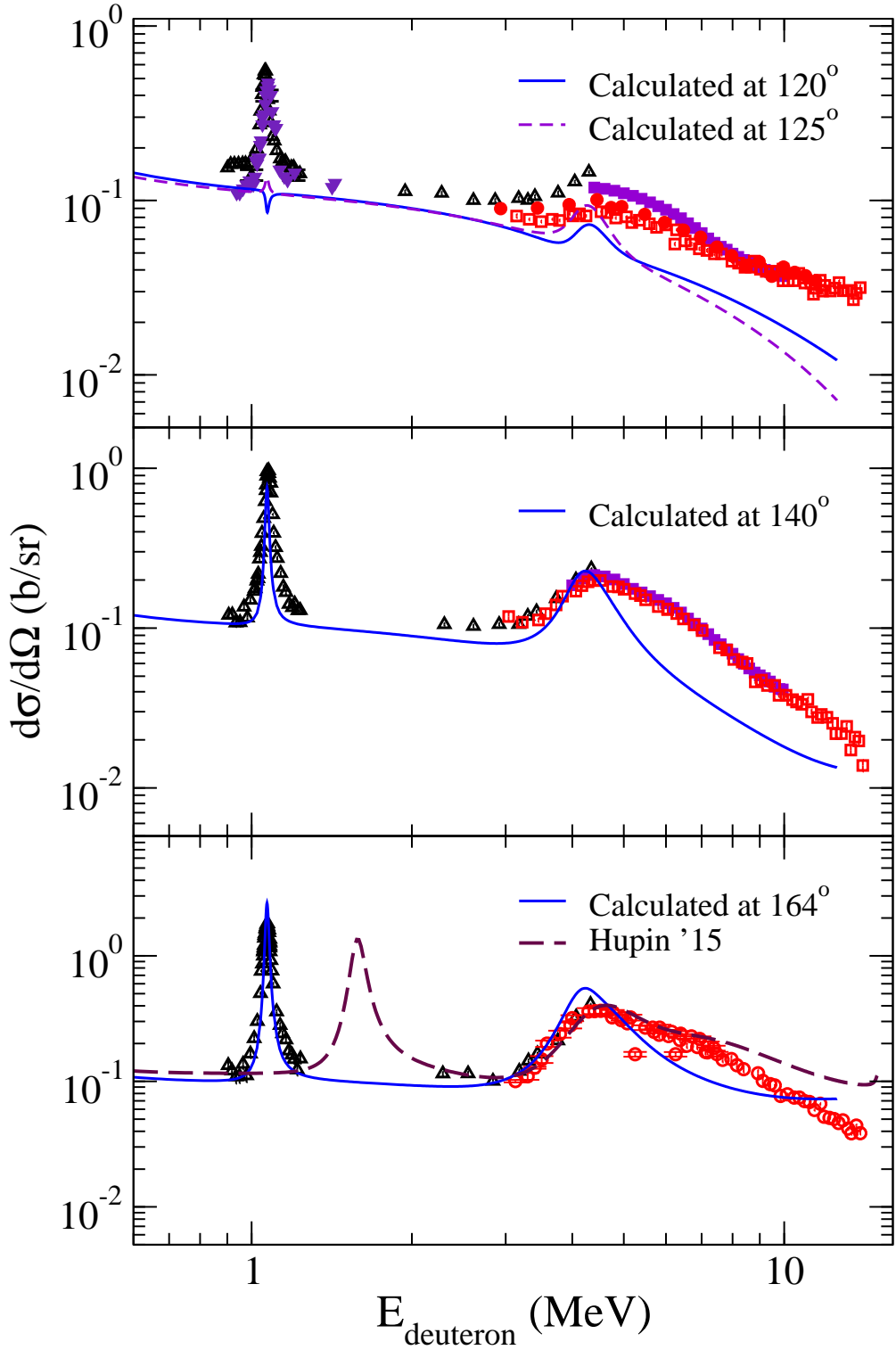}}
\end{center}
\caption{ \label{Fig4}(Color online.) Experimental elastic cross
  sections for $^2$H+$^4$He, at fixed angles, compared with
  the calculations of this work and Ref.~\cite{Hu15}.
  The data are from Refs.~\cite{La53,Ga55,Oh64,Se64,Ma68,Ba79}.}
\end{figure*}

Cross sections calculated at fixed scattering angles using the
associated $S$-matrices of Eq.~(\ref{multiS}) angles, are compared to
measured data in Figs.~\ref{Fig3} and \ref{Fig4}. The angles at which
calculations have been made are shown in each segment of these
figures. The data shown in these figures are taken from
Ref.~\cite{Se64} (filled circles) at 37.2, 50.0, 51.67, 90.0, and
120.0$^\circ$, from Ref.~\cite{Ba79} (open circles) at 38.75, 48.9,
90.0, and 125.0$^\circ$, from Ref.~\cite{Oh64} (filled squares) at
51.9, 90.0, 125.3, and 139.1$^\circ$, from Ref.~\cite{Ma68} (open
squares) at 50.36, 87.23, 120.1, 137.5, 163.0, and 164.5$^\circ$, from
Ref.~\cite{Ba79} (upside down triangles) at 38.75, 48.9, 90.0, and
125.0$^\circ$, from Ref.~\cite{La53} filled triangles) at 90.0 and
120.0$^\circ$, and from Ref.~\cite{Ga55} (open triangles) at 90.1,
125.2, 140.7, and 167.7$^\circ$. They are given in the segments in
which they are closest to the calculation angle.  All cross sections
are in centre-of-mass frame, and projectile energies are all in
laboratory frame with an $\alpha$-particle target. While the
calculation is defined with a deuterium target, the appropriate change
of frames has been performed.

In both Fig.~\ref{Fig3} and \ref{Fig4}, two calculated resonance
features are evident. They coincide with the first excited, isoscalar,
$3^+$, and the isoscalar $2^+$ states of ${}^{10}$Be.  In the middle
panel of Fig.~\ref{Fig3}, the locations of the experimentally known
and calculated states of ${}^{10}$Be are shown. In Fig.~\ref{Fig4},
wherein our results are compared with data taken at backward
scattering angles, to more clearly see the structures, the plots are
fully logarithmic.  Again the $3^+$ and $2^+$ resonances are most
evident and the calculated results for energies above $\sim$5~MeV are
too small, not revealing any resonance effect due to formation of the
isoscalar $1^+$ and of the isovector $2^+$ states.  Also shown in the
bottom panel is a second calculated result taken from
Ref.~\cite{Hu15}. Their model gives a better description of the data
in the 4 to 8~MeV region.
%%%%%%%%%%%%%%%%%%%%%%%%%%%%%%%%%%%%%%%%%%%%%%%%%%%%%%%%%%%%%%%%%%%%%%%%%%%%%%%
%%%%%%%%%%%%%%%%%%%%%%%%%%%%%%%%%%%%%%%%%%%%%%%%%%%%%%%%%%%%%%%%%%%%%%%%%%%%%%%
%%%%%%%%%%%%%%%%%%%%%%%%%%%%%%%%%%%%%%%%%%%%%%%%%%%%%%%%%%%%%%%%%%%%%%%%%%%%%%%
More specifically, the shape of the $3^+$ resonance is recreated well
at most scattering angles; centroids, widths and with reasonable
strengths.  The exception is the result for $\theta_{cm} = 125^\circ$
where, while the resonance effect is noted at the correct centroid
energy, the magnitude is too low. Off resonance, our calculated results
agree by and large with the available low energy data.  For the higher
energy region, the resonance feature due to formation of the $2^+$
state is well recreated at 50$^\circ$ and 164$^\circ$, and reasonably
well at some of the other angles. The non-resonant background
calculated at energies above this resonance usually is underestimated
and the $1^+$ resonance present in the data is not reflected in our
calculated results. This resonance was found by the six-body
calculation of $^2$H-$^2$He scattering by Ref.~\cite{Hu15}, however, and
so is a distinctive difference in the results of a more sophisticated
calculation than ours.

\begin{figure*}[htp]
\begin{center}
\scalebox{0.75}{\includegraphics*{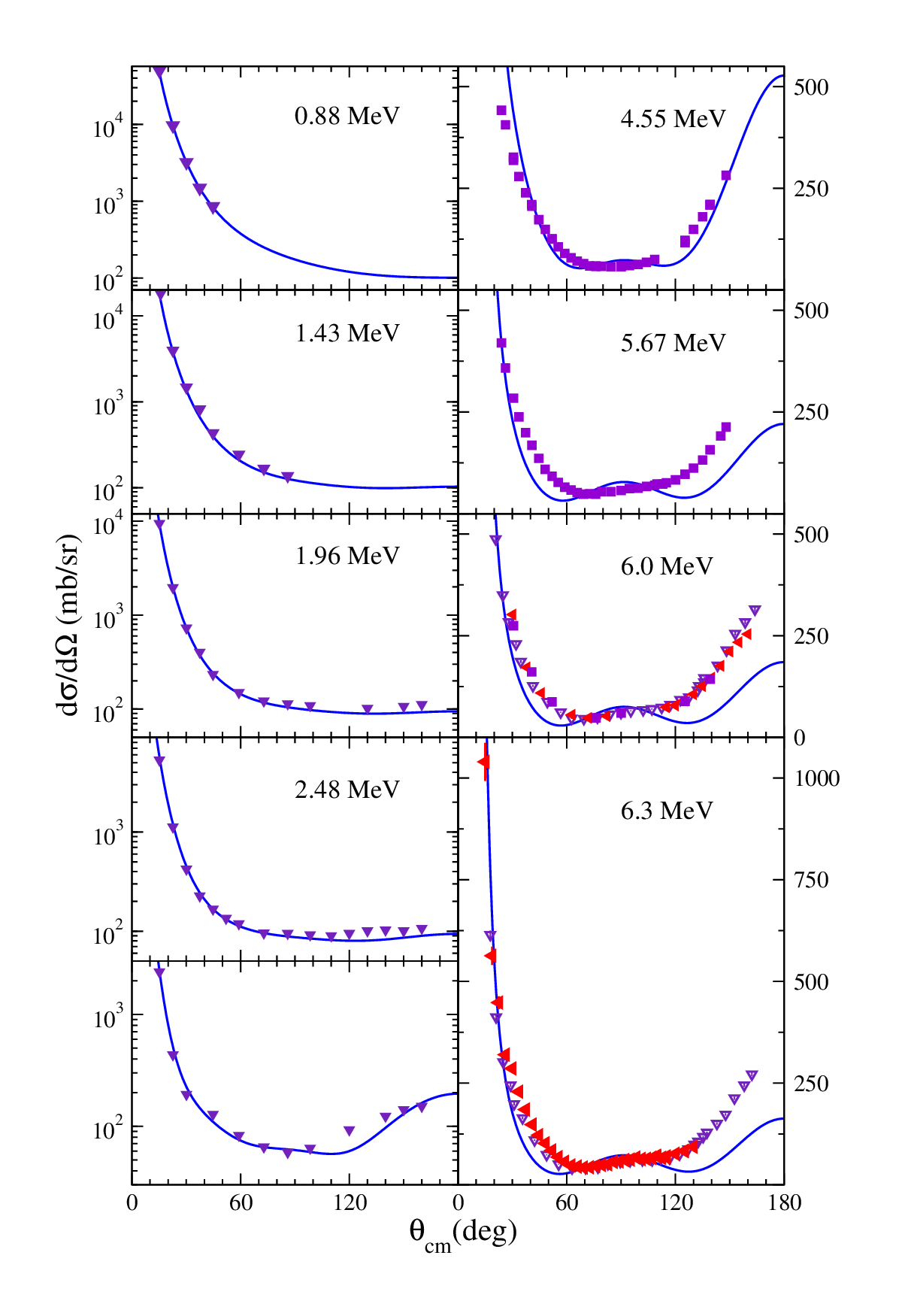}}
\end{center}
\caption{ \label{Fig5}(Color online.) Experimental elastic cross
  sections for $^2$H+$^4$He scattering, at fixed deuteron energies,
  compared with the calculations. The data are taken from Refs.~\cite{Bl49,St62,Oh64,Ha77}.}
\end{figure*}

\begin{figure}[htp]
\begin{center}
\scalebox{0.75}{\includegraphics*{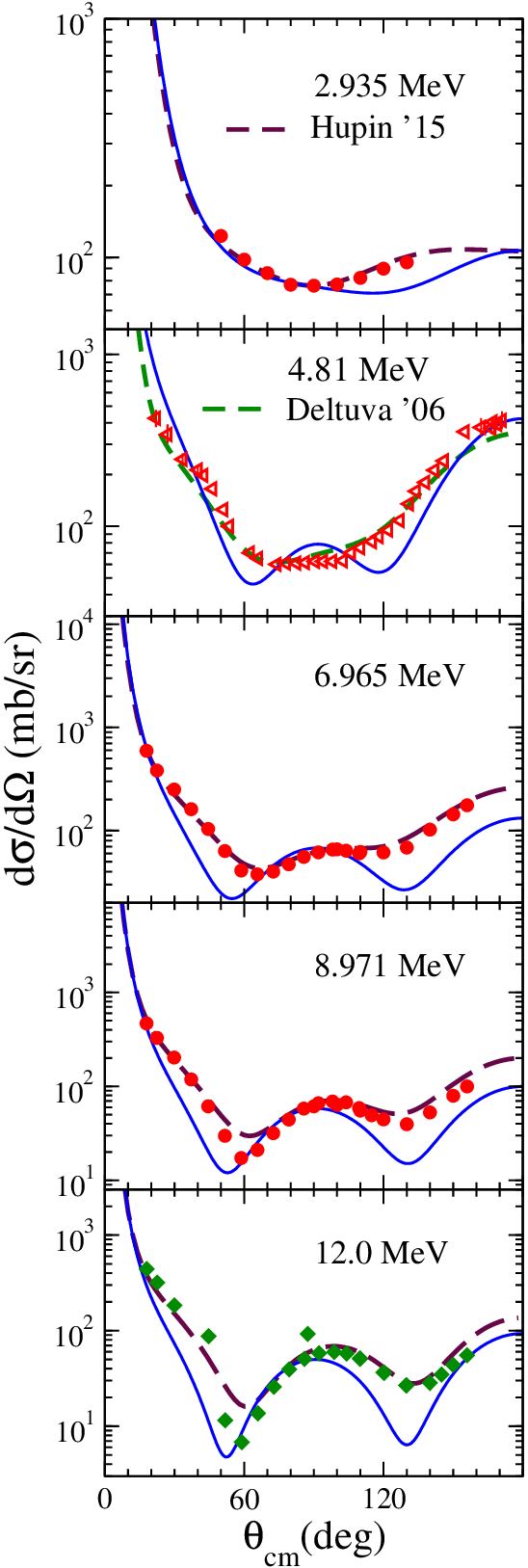}}
\end{center}
\caption{ \label{Fig6}(Color online.) Experimental elastic cross
  sections for $^2$H+$^4$He scattering, at fixed deuteron energies,
  compared with calculations of this work and
  Refs.~\cite{De06,Hu15}. The data are from
  Refs.~\cite{Se64,Je71,Br80}.}
\end{figure}

Cross sections calculated at fixed energies are compared to experiment
in Figs.~\ref{Fig5} and \ref{Fig6}. The former shows differential
cross sections for eight deuteron energies, ranging from 0.88 to
6.3~MeV. For clarity, the results and data in the left hand panel are
depicted semi-logarithmically, those in the right hand panel are shown
on linear scales.  In Fig.~\ref{Fig6} we examine five data sets, four
of which were also studied in Ref.~\cite{Hu15}, at 2.935, 6.695, 8.971
and 12~MeV, and the fifth that was studied in Ref.~\cite{De06}. The
notation is as given for Figs.~\ref{Fig3} and \ref{Fig4} with
additional data depicted as follows; Ref.~\cite{Bl49} (filled inverted
triangles), Ref.~\cite{Ha77} (open inverted triangles),
Ref.~\cite{St62} (left filled triangles), Ref.~\cite{Br80} (open left
triangles), and Ref.~\cite{Je71} (filled diamonds).
%%%%%%%%%%%%%%%%%%%%%%%%%%%%%%%%%%%%%%%%%%%%%%%%%%%%%%%%%%%%%%%%%%%%%%%%%%%%%%%
%%%%%%%%%%%%%%%%%%%%%%%%%%%%%%%%%%%%%%%%%%%%%%%%%%%%%%%%%%%%%%%%%%%%%%%%%%%%%%%
%%%%%%%%%%%%%%%%%%%%%%%%%%%%%%%%%%%%%%%%%%%%%%%%%%%%%%%%%%%%%%%%%%%%%%%%%%%%%%%

With some exceptions, this two-body calculation tends to reproduce the
small-angle scattering better than data at larger angles, matching
small angle data slightly better than the results given in
Ref.~\cite{Hu15}. However the results found in Ref.~\cite{Hu15} are superior
to ours at the large scattering angles.

For the lower set of energies, as shown in Fig.~\ref{Fig5}, our
calculated results agree quite well with the data, especially at the
four lowest energy values that span the region of the $3^+$ resonance.
The 4.6~MeV result, near the 2$^+$ resonance, is quite a good match to
data. Above this energy, where the 1$^+$ resonance is expected to
influence results, our results are poorer, as may be expected. In
general, at deuteron energies from 4.5 to 6.3~MeV, the calculated
cross sections have shapes more pronounced than in the data.

In Fig.~\ref{Fig6} we compare a select set of data and our results
with the differential cross sections given in Refs.~\cite{Hu15,De06}.
The latter results, shown by the dashed curves, are in excellent
agreement with the data at all of the selected energies.  Our results
are not in as good agreement, but the shapes and magnitudes of them
are acceptable in comparison with those revealed in the data.

\section{Conclusion}
\label{concl}

The methodology we have used enables all low excitation compound
system properties, spin-parities, energies and widths, extractable
from a specific Hamiltonian to be found.  With it, allowance can be
made for the effects of the Pauli principle in regards to assumed
occupancies of nucleon orbits in the target states. For single
channel problems such as those addressed herein, without such
accounting (via orthogonalizing pseudo-potentials), spurious states are
unique and orthogonal to those that are not. Thus, they can simply
be discarded.  With the cases studied, all spectral properties are
found by solving Lippmann-Schwinger equations. Resonance
properties are defined by the poles of the $T$-matrix associated with
the chosen Hamiltonian.

The first cases considered were $^7$Li and $^7$Be formed as the
clusters of $^4$He with $^3$H and $^3$He respectively.  As the
$\alpha$ break-up thresholds are 2.47 and 1.49~MeV respectively, states
above those energies were found that are resonances in the cluster
evaluations with widths that agree quite well with observation.  The
widths of resonance states are reaction specific but as only the
$^4$He break-up channels are relevant in the energy range considered
(the next threshold is 7.25~MeV for neutron emission from $^7$Li and
5.61~MeV for proton emission from $^7$Be), those widths then are also
the total widths.  The good agreement with experimental values is
evidence of the model's utility.

The next study made was that of the spectrum for $^8$Be formed as a
cluster of two $^4$He; a process at the heart of the so-called three
$\alpha$ formation of the Hoyle state in ${}^{12}$C in stellar
environs.  With our two-body approach, we find two low-excitation
resonance states in ${}^8$Be.  They are the ground state ($0^+$)
resonance having centroid and width energies of 0.092~MeV and 5 eV
[c/f experimental values\cite{Ti02} 0.092~MeV and 5.96 eV] and a first
excited ($2^+$) resonance state with centroid and width energies of
3.16~MeV and 1.11~MeV compared with experimental values of 3.03~MeV
and 1.51~MeV respectively.  Starting with this, we plan full
coupled-channel calculations of the $^4$He+$^8$Be cluster leading to
the Hoyle state.

We then considered $^6$Li as a $^2$H+$^4$He cluster. We considered the
two states of the $^2$H, the ground $^3$S$_1$ and the $^1$S$_0$,
as uncoupled states and solved two single channel LS equations
to obtain estimates of the isoscalar and isovector states in
the low-excitation spectrum of $^6$Li.
Four of the possible six states were found in good agreement with
the known values~\cite{Ti02}, with only the two highest ones,
the $2^+_2$ and $1^+_1$ differing by an MeV from the correct energies.

We have also made calculations of $^2$H+$^4$He scattering at low
energies, treating both as single bodies. It was found that this
approach recreates many of the features observed experimentally,
though some require a more sophisticated approach. The $^4$He ground
state was coupled to the $^2$H ground state treated as a pure
$^3$S$_1$ state, and separately to a $^1$S$_0$ resonance, to calculate
the spectrum of $^6$Li. Channels of the $^3$S$_1$ and $^1$S$_0$ states
were not coupled. All six known low-energy $^6$Li states were
recreated, with the first four very close to their known energies and
the two most energetic being found at energies that deviate from the
measured states by $\sim$1~MeV.  The $^2$H and $^4$He ground states
were coupled to calculate elastic scattering cross sections, and the
match to data was overall good. The observed $3^+$ and $2^+$
resonances were recreated, and had the correct shapes and reasonable
magnitudes at most angles. The non-resonant cross section was also
well reproduced. The observed $1^+$ resonance, however, was not
evident in calculated cross sections, though the state is found in the
calculated spectrum. Cross sections at fixed angles were good near the
two observed resonance energies, though in general results at low
angles were a better match to data than those at high angles.

A gauge invariant theory to evaluate capture cross sections using the
bound and continuum wave functions derivable from solutions of the
Lippmann-Schwinger equations has been developed (and used) for
${}^3$H+$^4$He system~\cite{Ca08}. Studies of the other cases
discussed herein, being important astrophysical quantities, are
planned for a future publication.

%%%%%%%%%%%%%%%%%%%%%%%%%%%%%%%%%%%%%%%%%%%%%%%%%%%%%%%%%
\begin{acknowledgments}
The authors gratefully acknowledge G. Hupin, S. Quaglioni,
P. Navr\'atil and A. Deltuva for providing numerical results of their
published cross sections.  This work is supported by the Australian
Research Council and U.S. National Science Foundation under Award
No. PHY-1415656.
\end{acknowledgments}
%%%%%%%%%%%%%%%%%%%%%%%%%%%%%%%%%%%%%%%%%%%%%

\bibliography{Fr17}

\end{document}